# Integration of a Predictive, Continuous Time Neural Network into Securities Market Trading Operations

**Christopher S Kirk**
*Forecast Limited, UK*
drcskirk@forecast.ltd.uk

**Abstract**

*This paper describes recent development and test implementation of a continuous time recurrent neural network that has been configured to predict rates of change in securities. It presents outcomes in the context of popular technical analysis indicators and highlights the potential impact of continuous predictive capability on securities market trading operations.*

## 1 Introduction and early development

This paper presents details of research and development progress regarding the systems integration of a neural network as a subsystem of a functional trading and execution system. This research is divided into two phases; the first phase of this study comprised the development of the core neural network algorithms and back-testing of the base system in an offline mode to determine suitability. The second phase is the conversion to a full online system with integration into active trading operations.

## 2 Overview

Neural networks can assist in resolving problems where solutions require adaptation to different circumstances and where development of an algorithm is complex or not possible. They are a non-classical use of computers that deploy a high number of connections and computational units in virtual-space and time which the algorithm can modify. In doing so they can take advantage of opportunities for massive parallelism and are said to 'learn'. This is in contrast to the classical use of computers that certainly operate very quickly, but only do precisely that which the programmer instructed and cannot 'learn'. In a manner similar to numerical methods it deploys numerical algorithms as 'learning algorithms'. These algorithms refine approximations through minimisation of error and successfully describe the behaviour of complex systems such as in the study of Computational Fluid Dynamics (Kirk and Mileham, 1999 and 2002).

The driving force for this project is the hypothesis that connectionist theory and its application to natural language processing, coupled with a concept for computation in continuous time  using soft constraints, can be applied to model and simulate the market price and rate of change of a security.  This hypothesis has been tested on the data feed of a number of different asset classes.  This is considered to be highly relevant to the integration and adoption in Securities Market Operations since different assets and indices display individual characteristics in price range, volatility and response to external influences as illustrated the exchange rates in Figure 1 below.  The requirement is for an adaptive algorithm that will continuously model satisfactorily, with good correlation, across a range of target problem domains such as these, without intervention.

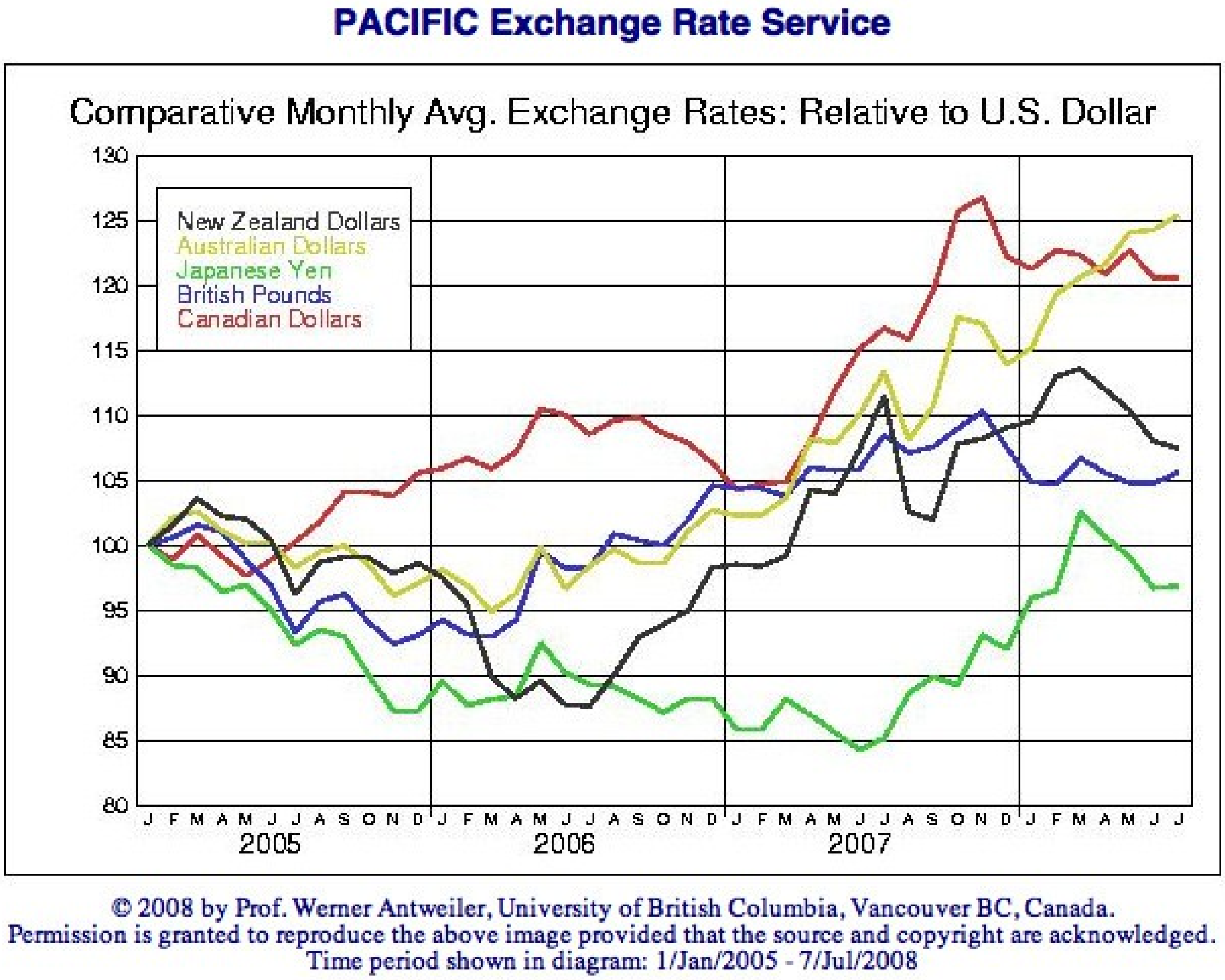

**Figure 1: A comparison of range differences across different exchange rates**

Further, the widespread use of automated order execution inhibits short-term opportunities due to programmed buying and selling at higher frequencies that are not able to be modeled effectively by the existing range of technical indicators. It is considered that a potential advantage is the ability to improve trading and market-making operations through an efficient, robust predictive capability to derive close approximations for rates of change, to exploit statistical arbitrage opportunities and to detect hidden, non-linear patterns.

## 3 Research and Development

### 3.1 Phase I

A conceptual model was developed that comprised data acquisition and pre-processing, a neural network providing a computational core with a post-processing execution subsystem. This was a 'classical' feedforward neural network (illustrated in Figure 2) that provided an offline test framework to assess the suitability of a connection-based network. It was configured to accept inputs from a number of pre-processed data feeds (such as intermarket indices and economic indicators). The output was a vector showing the predicted scope and direction of the target signal line (eg stock price) being evaluated.

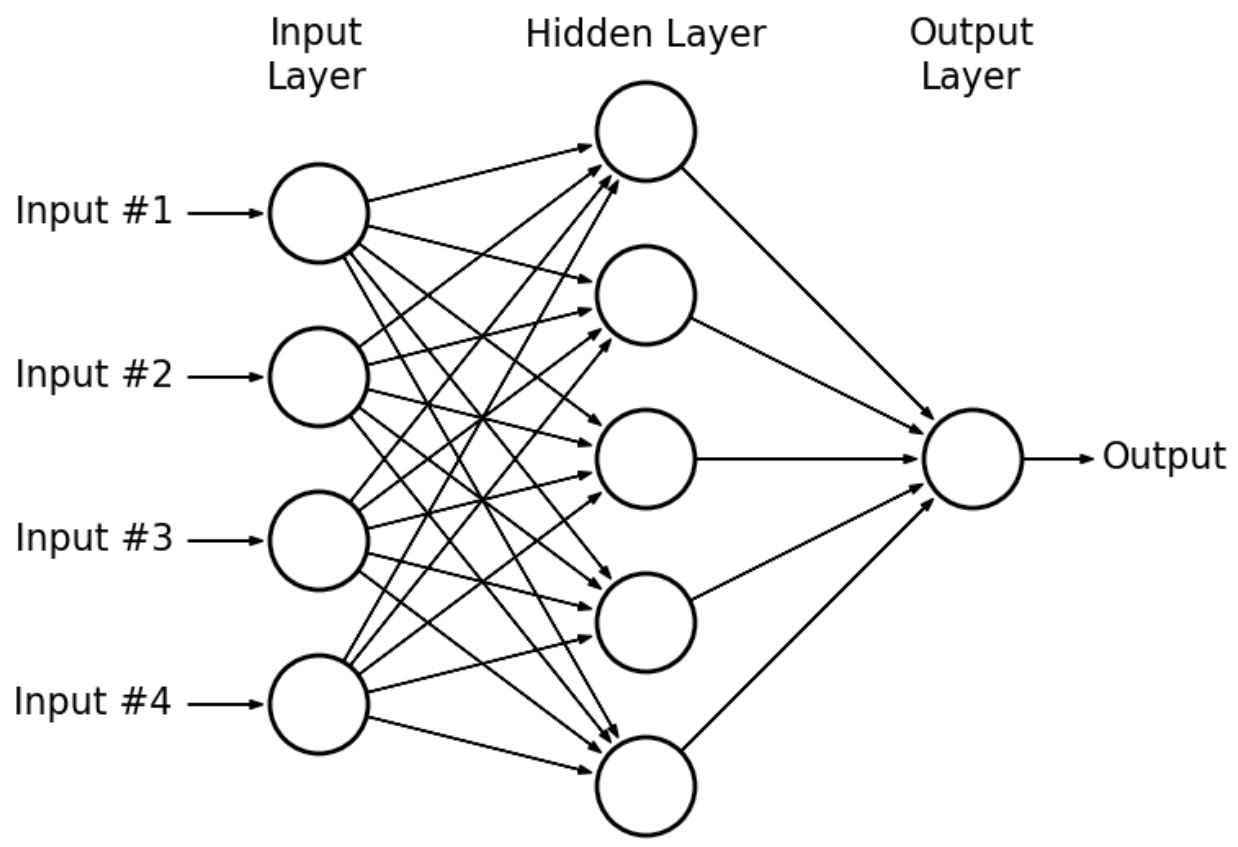

Figure 2: Conceptual framework in a typical neural network, (after Vanderplas, J, 2014, reprinted with kind permission)

These trials were successful and did provide the basis for use as a prediction model, but it was concluded that both accuracy and definition of the type of end-use application might be considered for further work. It is this further work that is the main basis for this paper.

The system described in this part one was largely developed for automated (computer to computer) trading only and had no graphical output. The system was therefore further refined later to show both graphical and textual output on a per-stock basis for the comparison of forecasts with actual price quotations. The back-testing of a continuously evolving dynamic, adaptable system is not trivial. However it was concluded that a back-test, run as an offline (batch-processed) evaluation of an imitation long/short portfolio of a 'basket of twenty stocks' selected daily from 150 UK equities, for a test period of two months would be a reasonable test. The holding period for each equity holding in this 'basket of stocks' was not fixed. Classification for entry into the imitation portfolio was on the basis of statistical assessment. Those yielding the highest Pearson-correlation during the training-process were selected. The back-test yielded the following results.

Table 1: Results of the backtest of an imitation portfolio of a 'basket of 20 stocks':

| Position in portfolio | No. of days in test | Initial capital | Valuation at 2 months | ROI | Peak Day | Valuation at Peak | Peak ROI |
|---|---|---|---|---|---|---|---|
| 0 – 10 | 41 | 1000 | 1428 | 42.80% | 32 | 1490 | 49.00% |
| 11 – 20 | 41 | 1000 | 1462 | 46.20% | 38 | 1471 | 47.10% |

The neural network is by design, dynamic and self-correcting and the batch processing was considered reasonable for an assessment of suitability. The backtest results in Table 1, illustrate resilience and reasonable performance in an average market. Notably, each time the 'basket of stocks' was selected it could potentially contain completely different stocks from the previous run. This might not correspond to an optimal trading strategy. It is also not unreasonable to see that these might not be performers with the best return or risk/return ratio, but it is reasonable to expect that this system performance will be replicated for other back-test periods. Of course, extending the subset of equities to alternative sectors or asset classes would similarly change the potential structure and return, as would the conversion from mathematical prediction to traded gains after costs by active trading. The results in Table 1 are theoretical limits only, shown gross and do not account for the costs of trading.

### 3.2 Phase II

The neural network potential and back-test was considered satisfactory and it was determined that this should be extended from batch processing (offline) to continuous operations (online), to advance the provision of solutions to meet the needs of active traders and other professionals operating in higher-frequency timeframes.

The network elaborated in Phase I, was a 'classical' feedforward backpropagation methodology deploying a gradient descent procedure to evaluate (and minimise) the error derivatives as a single network. This was further developed to refine the calculation of the error gradient (in a manner similar to linear discretisation in numerical analysis), by dividing the problem domain into a series of internal states between each canonical time-step. The use of the back-propagation through time methodologies (McClelland, 2013) approximates and refines the activation and error calculations during each time-step as internal states and is similar to a computational unfolding of a single network into a number of sub-systems or networks and is illustrated in Figure 3.

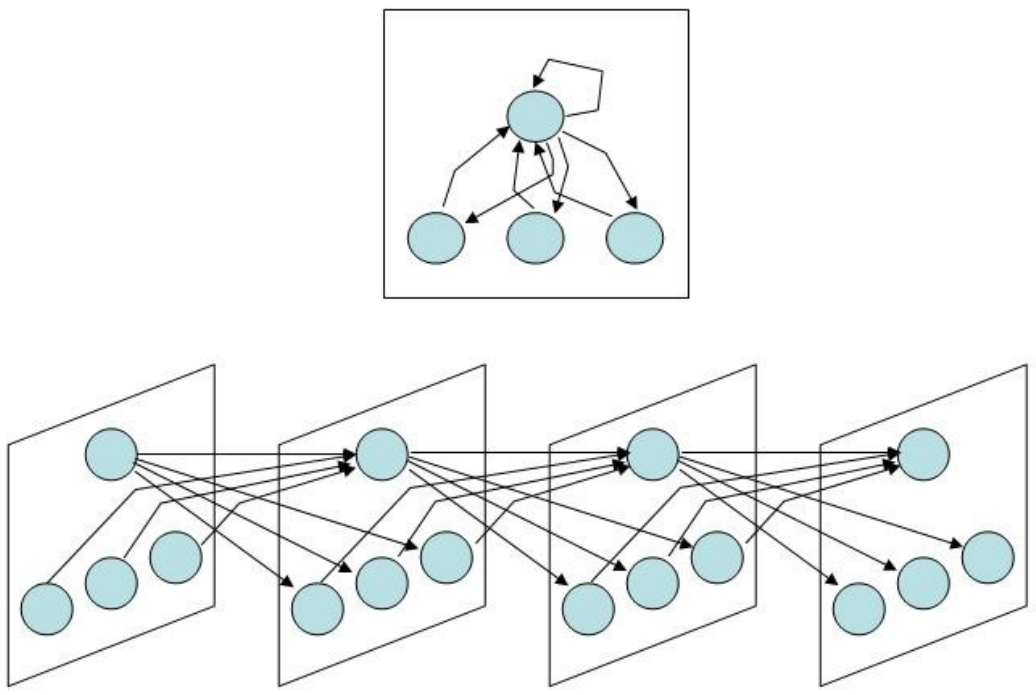

Figure 3: The time domain discretised into a computational mesh of time-steps (dt). *(Reprinted with permission from Figure 8.2 of McClelland, J. L. (2013). Explorations in Parallel Distributed Processing: A Handbook of Models, Programs, and Exercises, [http://www.stanford.edu/group/pdplab/pdphandbook/](http://www.stanford.edu/group/pdplab/pdphandbook/). Copyright James L. McClelland.*

This, (BPTT) was further elaborated to allow the new network to evolve with a continuous-time capability by replacing the discrete internalised states with a set of coupled differential equations after the work of (Pearlmutter 1988 and 1989, Williams and Zipser, 1989). During the simulation, the discretisation of the temporal domain into a computational mesh (on which the PDEs are subsequently discretised) and the algorithm computes its gradient in weight space, adjusts the weights and the resultant outputs then form the input for the next time step. These modifications were potentially computationally expensive and so they were rationalised after the work of (Williams and Peng, 1990 and Williams and Zipser, 1995). The final algorithm now computes simultaneously across multiple processor nodes and the error checking routines are continually summed rather than at the end of each epoch. Subject to the supply of inbound data, this allows the network to function continuously in an online mode.

To further optimise the network, batch-learning was replaced with a stochastic algorithm after the work of LeCun, 1989 and the learning-rate was automated more elegantly which resulted in a much faster solution.

## 4 Application to trading operations

### 4.1 Market data feed

Market data feed, charting and execution (through a broker) are typically embodied in integrated software packages or electronic trading platform. Large firms and institutions might also integrate risk, compliance and P&L/reporting systems with specialist feeds pertaining to the derivatives and synthetic products that they create. The supply of external market data is tiered by frequency to provide access to low latency feeds and to provide additional information other than the bid/ask price, such as order flow and market heuristics. A continuous-time neural network of the type described herein would function simultaneously with the same data feed and electronic trading platform in order not to limit functionality.

### 4.2 Technical Analysis Indicators

Classical technical analysis dating back to at least 1700, (Neely and Weller, 2011), is the statistical evaluation of past price and volume data and is used as the basis for indicator-based signal generation and for charting (Bulkowski, 2000), (pattern matching success is highly dependent on the experience of the user). Whilst styles of trading might be very different, technical analysis indicators are intended for use in both the trending phase or the non-trending (sideways/range-bound) phases and are unaffected by style or strategy. As an addition to these indicators, a virtual structure, (support and resistance), can be charted by using local minima/maxima price values where previous trends were interrupted or at previous session end-values or volatility ranges. Additional chart subdivisions (Murphy, 1999), may be based on a Fibonacci number sequence referencing previous price ranges or Elliott waves. Technical analysis indicators, were derived before the advent of modern computers to quantify a number of price relationships. They were usually formulated with a fixed temporal window, (Murphy, 1999) that was matched empirically to the price cycle of the asset being analysed. Despite being intended for specific asset classes or market phases, many of these indicators are deployed out of their original scope and can therefore provide contradictory results when compared with others.

Whilst these indicators describe only a snapshot of a non-linear dynamical system that itself is evolving, one profound value of charting technical analysis indicators is that the user can see a change visually and moreover a change relative to previous price history and this can be interpreted mentally by the user to imply a future state. Despite their shortcomings they appear to be in widespread use (Neely and Weller, 2011) and therefore in this paper, output from the neural network is considered best described relative to established technical analysis indicators rather than an academic discussion of vectors and their modeling constructs.

Typical, classical technical indicators are summarised in Table 2.

Table 2: Classes of Technical Indicators

| Type | Name | Timing | Market phase |
|---|---|---|---|
| Trend-following | Moving average, MACD | Lagging | Trend |
| Oscillators | RSI,Stochastic, CCI | Visually Leading | Range-bound |
| Volatility | Ave Range, Bollinger, Keltner | Lagging | Trend/Range-bound |
| Volume | Money Flow, On-balance | Lagging | Trend/Range-bound |
| Structure | Fibonacci Numbers, Support and Resistance | Implied Leading | Trend/Range-bound |

If outputs from the neural network are to be considered in the context of existing technical analysis indicators then it is necessary to attempt to rationalise those indicators and adapt them to the obvious major impact of the variations in the length of cycle or wave embodied in the securities price.

### 4.3 Adaptation using Wavelet Transforms

In other work, the wavelet-based time-frequency analysis of an evolving data feed (Raihan et al., 2005) enabled the evaluation of both the frequency and the timing of asset price cycles and the mapping of the rate of change as each cycle shifts in time. This was applied to evaluate the length of the wave cycle and from it form a basis for adapting the indicators from fixed-period indicators to adaptive-period indicators.

Cycle-analysis of each asset is vital to ensure that the correct market phase (trend or range-bound) model is being used. Deployment of adaptive technical indicators complements existing control systems (such as a Fuzzy Inference System, FIS) with little additional code revision cost. Figure 4, illustrates the predictive output (5-min timeframe) from the neural network compared with the actual price history on which it was validated and trained. Each chart includes the system calculation of Pearson-correlation for the two traces. The chart is complemented with adaptive indicators (MACD and Stochastic K/D) with a simulation of the wavelet transform used to evaluate the wave cycle length. As a guide, note that -10 on the x-axis equates to '50 minutes ago' from the current time (zero) when evaluated using this 5-minute timeframe. In this simulation the values were predicted 10 periods ahead, (8 periods plotted) and the technical analysis indicators used these values as the basis for their calculation.

It is considered that it is with this combination chart that the transition to the use of neural networks is likely to be of most use to traders. Most platforms currently only have provision to chart historical values (since future prices cannot be part of current data feeds) though modification is considered trivial.

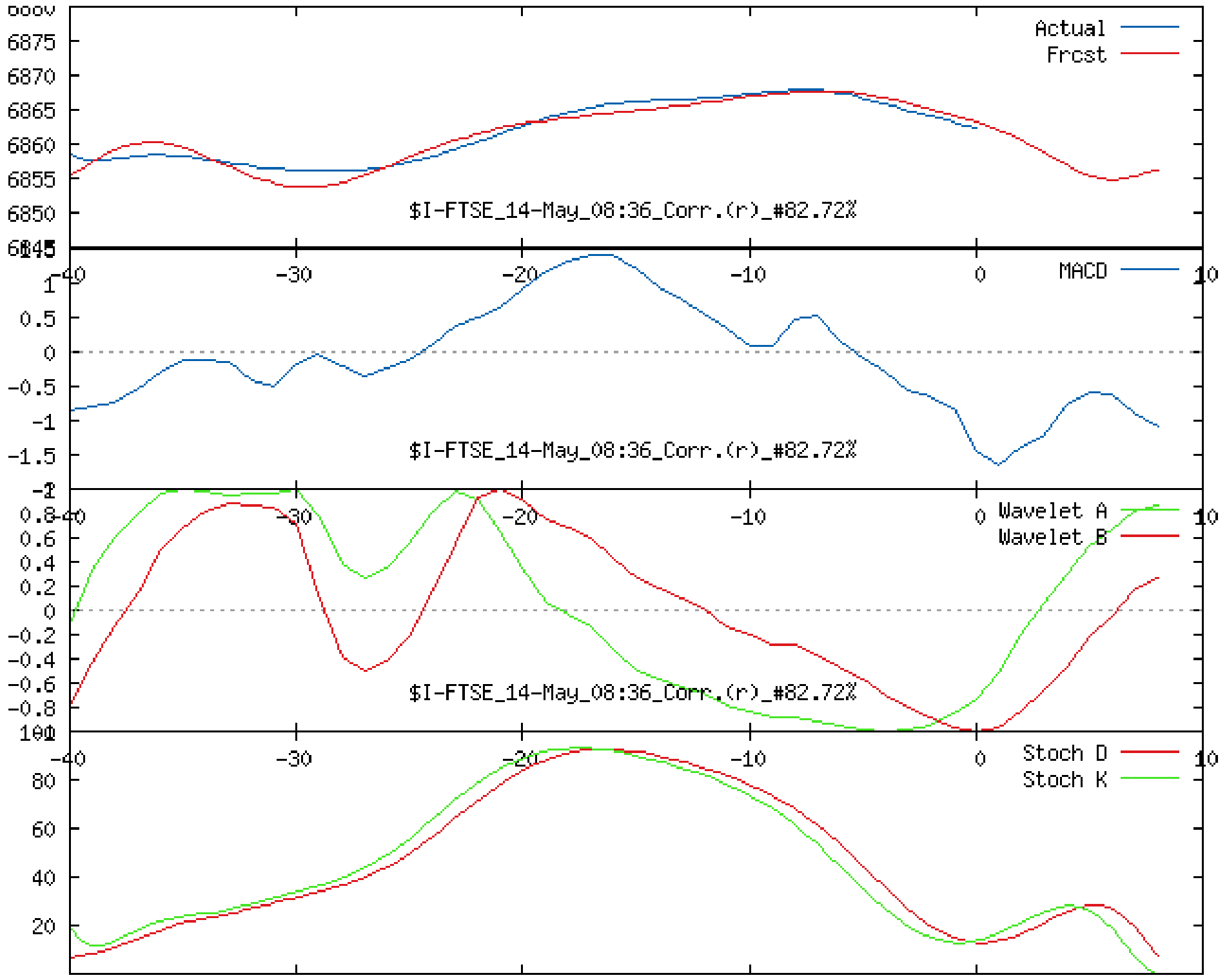

Figure 4: Example chart produced from the continuous output for the FTSE 100 Index. Actual price (blue) with prediction (red) with supporting adaptive technical analysis indicators, MACD, Wavelet growth, Stochastic K & D

**4.4 Continuous output – continuous trading**

A major driver for using automated strategies is to implement secure, reliable, (rule-based) trading procedures within a risk-controlled regime. It follows that progression into semi-autonomous trading functionality requires a step-change in the use of pre-trade modeling and integration of that modeling into control systems.

In utilising the continuous output from a simulation such as described herein, there may be a need to modify investment and trading strategies. As an example, in parallel work, (Panella, 2008) the introduction of machine learning systems and artificial intelligence methodologies to improve the surveillance, target detection and tracking systems of an unmanned aerial vehicle (*UAV*) also requires adaptation of automatic systems and in some cases replacement by autonomous systems. Thus, automated trading systems may need to be replaced by autonomous or semi-autonomous systems that allow them to 'learn'. The system can then adapt and act more efficiently in response to unanticipated events and reduce risk accordingly. This would turn the trading operation into a highly objective function rather than a highly subjective one. During the integration phase there will be a need to with further adaptation in time.

Since levels of autonomy might be classified, [low, medium, high] according to function, (Panella, 2008) this can be translated in securities market terms to address the preservation of capital and liquidity, the extent of capital employed, the overall risk exposure and ultimately a more defined approach for monitoring not just the current asset value, but the evolving prediction delivered by the neural network on a continuous basis, perhaps by creating a hybrid system using a form of mean-reversion strategy where a moving average of the output is an advanced mean, or by considering continuous trading using a scaling strategy.

## 5 Systems Integration

### 5.1 Parallel operations on a Cluster

A downside noted by [Williams Zipser 1989a] was that some adaptations of a neural network are computationally expensive and this implementation is no exception. Therefore, to complete the test phases a heterogeneous cluster of four, quad-core computers was built. This successfully proved the parallelisation of the code and the potential for scaleability as a distributed system. Costs and commissioning of a heterogeneous cluster compare favourably with supercomputers, notably where decommissioning is required, as the cluster can be decomposed into individual units that can be replaced as technology improves, without impacting on operations.

### 5.2 Pre-processing

A number of policies to decide the operational bounds for the system need to be formulated prior to submission of a continuous data feed for simulation. In particular the treatment of temporal discontinuities present in historical data such as the use of pre-trade and post-trade transactions, the impact of gap-openings where cash markets are modeled, the relevance of overnight transactions in rolling derivatives markets and foreign-exchange session-changeovers.

Market data feed can be noisy in that it can contain unwanted data which, if left in place would require unnecessary computation to model an extraordinary event and limit the ability of the model to simulate events realistically. It is therefore necessary to create a policy for the removal of outliers. These are sometimes referred to as 'Catastrophic Events' and can be 'fat-fingers', 'flash-crashes', reaction to news, or other excessively volatile price variances of very short duration. Market data is supplied by the vendor at various frequencies depending on contract and routing capability, ranging from the streaming of real-time data (stock market tick data), to a delayed feed where extraction files are pulled at pre-determined time steps of say, 1 minute, 3 minutes, 5 minutes or 15 minutes. Some history for real-time data feeds can be supplied but the period is often limited. A price and volume data feed was used in this project but additional feeds may be aggregated to include, for example, variance, vwap and order information.

## 6 Conclusion and Further work

In this paper we have presented an example of deep learning, namely the integration of a predictive, continuous time recurrent neural network into trading and risk assessment operations. During application within a trading environment the potential need to adapt technical analysis indicators that their use might be continued and integrated into a hybrid system was highlighted. It has been shown that this architecture can be directly applied to model pre-processed data feeds of different asset classes but that the development of new trading strategies to make best use of continuous predictive output might be necessary.

This work is to be extended to fully integrate feeder networks to create a cluster of networks to simulate fully autonomous trading and risk assessment operations.

**Acknowledgements**
The author expresses thanks to the Directors of Forecast Ltd and Market Forecast Ltd for the access to intellectual property to enable the realisation of this paper.